\documentclass[journal]{IEEEtran}
\usepackage{epsfig}
\usepackage{cite}
\usepackage{graphicx}
\usepackage{subfig}
\usepackage{amsmath}
\usepackage{amssymb}
\usepackage{verbatim}
\usepackage{algorithm}
\usepackage{algorithmic}
\usepackage{booktabs}
\usepackage{multirow}
\usepackage{mathtools}
\usepackage{breqn}
\usepackage{color}
\usepackage{url}

\ifCLASSINFOpdf
\else
\fi
\begin{document}
%
\title{Semi-Supervised Segmentation of Radiation- Induced Pulmonary Fibrosis from Lung CT Scans with Multi-Scale \textcolor{black}{Guided Dense Attention}}
%
%
%

\author{Guotai~Wang, 
	    Shuwei Zhai,
	    Giovanni Lasio,
	    Baoshe Zhang,
	    Byong Yi,
	    Shifeng Chen,
	    Thomas J. Macvittie,
	    Dimitris Metaxas,
	    Jinghao Zhou,
        and Shaoting Zhang 
\thanks{This work was supported by the National Natural Science Foundations of China [81771921, 61901084] funding. (Corresponding authors: Guotai Wang; Shaoting Zhang)}
\thanks{G. Wang, S. Zhai and S. Zhang are with the School of Mechanical and Electrical Engineering, University of Electronic Science and Technology of China, Chengdu, 611731, China. 
 E-mail: \{guotai.wang, zhangshaoting\}@uestc.edu.cn}
\thanks{G. Lasio, B. Zhang, B. Yi, S. Chen, T. J. Macvittie and J. Zhou are with Department of Radiation Oncology, 	University of Maryland School of Medicine, Baltimore MD 21201, USA.}
\thanks{D. Metaxas is with Department of Computer Science, Rutgers, the State University of New Jersey, Piscataway, NJ 08854, USA.}
\thanks{This work has been submitted to the IEEE for possible publication.
Copyright may be transferred without notice, after which this version may
no longer be accessible.}
}

\maketitle

\begin{abstract}
Computed Tomography (CT) plays an important role in monitoring  radiation-induced Pulmonary Fibrosis (PF),  where accurate segmentation of the PF lesions is highly desired for diagnosis and treatment follow-up. However, the task is challenged by ambiguous boundary, irregular shape, various position and size of the lesions, as well as the difficulty in acquiring a large set of annotated volumetric images for training. To overcome these problems, we propose a novel convolutional neural network called PF-Net and incorporate it into a semi-supervised learning framework based on Iterative Confidence-based Refinement And Weighting of pseudo Labels (I-CRAWL). Our PF-Net combines 2D and 3D convolutions to deal with CT volumes with large inter-slice spacing, and uses multi-scale guided dense attention to segment complex PF lesions. For semi-supervised learning, \textcolor{black}{our I-CRAWL employs pixel-level uncertainty-based confidence-aware refinement to improve the accuracy of pseudo labels of unannotated images, and uses image-level uncertainty for confidence-based image weighting to suppress low-quality pseudo labels in an iterative training process.} Extensive experiments with CT scans of Rhesus Macaques with radiation-induced PF showed that: 1) PF-Net achieved higher segmentation accuracy than existing 2D, 3D and 2.5D neural networks, and 2) I-CRAWL outperformed state-of-the-art semi-supervised learning methods for the PF lesion segmentation task. Our method has a potential to improve the diagnosis of PF and clinical assessment of side effects of radiotherapy for lung cancers. 

\end{abstract}


%
\IEEEpeerreviewmaketitle

\section{Introduction}
%
%
%
%
\IEEEPARstart{L}{ung} cancer is the leading cause of cancer deaths around the world. In 2018, the global number of new cases and deaths were 2.09 million and 1.76 million, respectively~\cite{Bray2018}. 
Radiation therapy is one of the most effective and widely used treatment for cancer patients~\cite{Baskar2012}. With the development of such treatment technologies, lung cancer death rate dropped 45\% from 1990 to 2015 among men, and 19\% from 2002 to 2015 among women~\cite{Siegel2018}.   However, about half of all cancer patients who receive radiation therapy during their course of illness will suffer from Radiation-Induced Injuries (RII) to the hematopoietic tissue, skin, lung and gastrointestinal (GI) systems~\cite{Baskar2012}. To prevent, mitigate or treat the RII plays an important role in improving the quality of radiation therapies. 

For lung cancer, the most common RII is the radiation-induced Pulmonary Fibrosis (PF)~\cite{Wilson2009}, i.e., inflammation and subsequent scarring of lung tissues caused by radiation, which could lead to breathing problems due to lung damage and even lung failure~\cite{Giuranno2019}. Observation and assessment of the PF lesions using Computed Tomography (CT) imaging is critical for diagnosis and treatment follow-up of this disease~\cite{Christe2019}. For an accurate and quantitative measurement of PF, it is desirable to segment the PF lesions from 3D CT scans. 
The segmentation results can provide detailed spatial distribution and accurate volumetric measurement of the lesions, which is important for treatment decision making, PF progress modeling, treatment effect assessment  and prognosis prediction. 

\begin{figure*}[t]
	\centering 
	\includegraphics[width=0.9\textwidth]{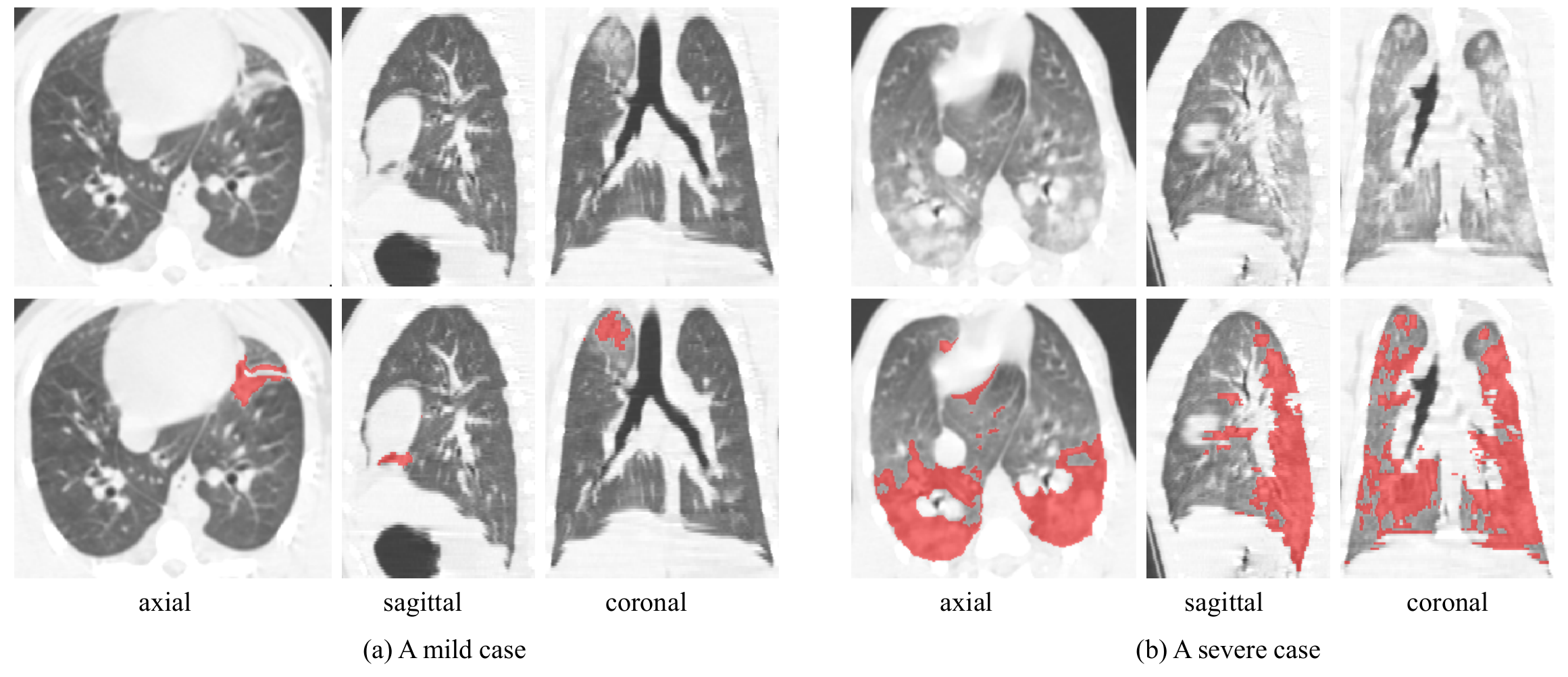}
	\caption{
	Examples of radiation-induced Pulmonary Fibrosis (PF) of the Rhesus Macaque. The first row shows lung CT images, and the second row shows manual segmentation results of PF lesions. Note that the ambiguous boundary, irregular shape and various size and position make the segmentation task challenging. }
	\label{fig:image_example}
\end{figure*}
As manual segmentation of lesions from 3D images is time-consuming, labor-intensive and faced with inter- and intra-observer varabilities,  automatic PF lesion segmentation from CT images is highly desirable~\cite{VanRikxoort2013}. 
However, this is challenging due to several reasons. Firstly, with different severity of the disease, PF lesions have a large variation of size and shape. A small lesion may only contain few pixels, while a large lesion can occupy a lung segment. The irregular shapes make it difficult to use a statistical model for the segmentation task~\cite{Wang2015c}. Secondly, the lesions have a complex spatial distribution, and can be scattered in different segments of the lung. Thirdly, PF lesions  often adhere to lung structures including vessels, airways and the pleura, and other lesions like lung nodules and pneumonia lesions with similar appearance may also exist. These factors, alongside with the low contrast of soft tissues in CT images, make it hard to delineate the boundary of PF lesions. Fig.~\ref{fig:image_example} shows two examples of lung CT scans with PF, and it demonstrates the difficulties of accurate segmentation.

Recently, Convolutional Neural Networks (CNNs) have been increasingly used for automatic medical image segmentation~\cite{Shen2017}. By automatically learning features from a large set of annotated images, they have outperformed most traditional segmentation methods using hand-crafted features~\cite{VanRikxoort2013}, such as for recognition of lung nodules~\cite{Xie2019,Wang2017e}, 
lung lobes~\cite{Xie2020}  and COVID-19 infection lesions~\cite{Fan2020,Wang2020c}. 
However, to the best of our knowledge, CNNs for PF lesion segmentation have rarely been investigated so far.

Besides the above challenges, existing CNNs may obtain suboptimal performance for the PF lesion segmentation task due to the following reasons. First, lung CT images usually have anisotropic 3D resolutions with high inter-slice spacing and low intra-slice spacing. Existing CNNs using pure 2D convolutions or pure 3D convolutions have limited ability to learn effective 3D features from such images, as 2D CNNs~\cite{Ronneberger2015,Zhou2018,Roy2019,Wang2015c} can only learn intra-slice features, and most 3D networks~\cite{Abdulkadir2016,Milletari2016,Liwenqi2017,Huang2019} are designed with isotropic receptive field in terms of voxels. When dealing with 3D images with large inter-slice spacing, they have an imbalanced physical receptive field (in terms of mm) along each axis, i.e., the physical receptive field in the through-plane direction is much larger than that in in-plane directions, which may limit effective learning of 3D features. 
Second, existing CNNs often use position invariant convolutions without spatial awareness, which makes it difficult to handle objects with various position and size. Attention mechanisms have recently been proposed to improve the spatial awareness~\cite{Oktay2018,Roy2019, Wang2017c}, but their obtained spatial attention does not match the target region well, and their  performance on PF lesions has not been investigated.  

What's more, current success of deep learning methods for segmentation relies highly on a large set of annotated images for training~\cite{Shen2017}. For 3D medical images, acquiring pixel-level annotations in segmentation tasks is extremely time-consuming and difficult, as accurate annotations could be only provided by experts with domain knowledge~\cite{Tajbakhsh2019}. For the PF lesion segmentation task, annotation of a CT volume could take several hours, and the complex shape and appearance of PF lesions further increase the efforts and time needed for annotation, which makes it difficult to annotate a large set of 3D pulmonary CT scans for training.

To deal with these problems,  we propose a novel semi-supervised framework with a novel 2.5D CNN based on multi-scale attention for the segmentation of PF lesions from CT scans with large inter-slice spacing. The contribution is three-fold. First, we propose a novel network for PF lesion segmentation (i.e., PF-Net), which employs multi-scale guided dense attention to deal with lesions with various size and position, and combines 2D and 3D convolutions to achieve balanced physical receptive field along different axes to better learn 3D features from medical images with anisotropic resolution.
Second, a novel semi-supervised learning framework using Iterative Confidence-based Refinement And Weighting of pseudo Labels (I-CRAWL) is proposed, \textcolor{black}{where uncertainty estimation is employed to assess the quality of pseudo labels of unannotated images in both pixel level and image level. We propose Confidence-Aware Refinement (CAR) based on pixel-level uncertainty to refine pseudo labels, and introduce confidence-based image weighting according to image-level uncertainty to suppress low-quality pseudo labels.}  Thirdly, we apply our proposed method to radiation-induced PF lesion segmentation from CT scans, and extensive experimental results show that our method outperformed state-of-the-art semi-supervised methods and existing 2D, 3D and 2.5D CNNs for segmentation. As far as we know, this is the first work on PF lesion segmentation based on deep learning, and our method has a potential to reduce the annotation burden for large-scale 3D image datasets in the development of automatic segmentation models with high performance.

\section{Related Works}
\subsection{CNNs for Medical Image Segmentation}
CNNs have achieved state-of-the-art performance for many medical image segmentation tasks~\cite{Shen2017}. Most widely used segmentation CNNs are inspired by U-Net~\cite{Ronneberger2015}, which is based on an encoder-decoder structure to learn features at multiple scales. UNet++~\cite{Zhou2018} extends U-Net with a series of nested, dense skip pathways for a higher performance. Attention U-Net~\cite{Oktay2018} introduced an attention gate using high-level features to calibrate low-level features. Spatial and channel ``Squeeze and Excitation" (scSE)~\cite{Roy2019} enables a 2D network to focus on the most relevant features for better performance. 

Typical networks for segmentation of 3D volumes include 3D U-Net~\cite{Abdulkadir2016}, V-Net~\cite{Milletari2016} and HighRes3DNet~\cite{Liwenqi2017}. 
They assume that the input volume has an isotropic 3D resolution to learn 3D features, and are not suitable for images with large inter-slice spacing. To better deal with such images, a 3D anisotropic hybrid network that uses a pre-trained 2D encoder and a decoder with anisotropic convolutions was proposed in~\cite{Liu2018c}. Jia et al.~\cite{Jia2020} designed a pyramid anisotropic CNN based on decomposition of 3D convolutions. The nnU-Net~\cite{Isensee2021} automatically configures network structures and training strategies, where different types of convolution kernels can be adaptively combined for a given dataset. In~\cite{Lei2019}, a lightweight CNN combining inter-slice and intra-slice convolutions was proposed to for segmentation of CT images. In~\cite{Wang2019b}, 2D and 3D convolutions were combined in a single network for Vestibular Schwannoma segmentation from images with anisotropic resolution. However, these networks have limited ability to segment lesions with various size and position. 
  
\subsection{Segmentation of Lung CT Images}
CNNs have been widely used for segmentation of lung structures from CT images~\cite{VanRikxoort2013}. 
In~\cite{Xie2020}, cascaded CNNs with non-local modules~\cite{Wang2017c} were proposed to leverage structured relationships for pulmonary lobe segmentation. In~\cite{Nadeem2020}, a CNN was combined with freeze-and-grow propagation for airway segmentation.  For lung lesions, a central focused CNN~\cite{Wang2017e} was proposed  to segment lung nodules from heterogeneous CT images, and Fan et al.~\cite{Fan2020} employed reverse attention and edge attention to segment COVID-19 lung infection. Wang et al.~\cite{Wang2020c} developed a noise-robust framework for automatic segmentation of COVID-19 pneumonia lesions by learning from non-expert annotations. Despite the large amount of works on lung structure and lesion segmentation so far, there is a lack of deep learning models for the challenging task of radiation-induced pulmonary fibrosis segmentation. 

\subsection{Semi-Supervised Learning}
To reduce the burden for annotation, semi-supervised learning methods have been increasingly employed for medical image segmentation by using a limited number of annotated images and a large amount of unannotated images~\cite{Tajbakhsh2019}. Existing semi-supervised methods mainly have two categories. The first category is based on pseudo labels~\cite{Lee2013,Bai2017a,Fan2020}, where a model trained with annotated images obtains pseudo labels for unannotaed images that are then used to update the segmentation model. Lee~\cite{Lee2013} used such a strategy for classification problems, and Bai et al.~\cite{Bai2017a} updated the pseudo segmentation labels and network parameters alternatively and used Conditional Random Field (CRF) to refine the pseudo labels. Fan et al.~\cite{Fan2020} progressively enlarged the training set with unlabeled data and their pseudo labels for learning. However, this method ignores the quality of pseudo labels, which may limit the performance of the learned model. 

The second category is to learn from annotated and unannotated images simultaneously, and they often consist of a supervised loss function for annotated images and an unsupervised regularization loss function for all the images. The regularization can be based on teacher-student consistency~\cite{Cui2019,Yu2019}, transformation consistency~\cite{Li2020}, multi-view consistency~\cite{Xia2020} and reconstruction-based auxiliary task~\cite{Chen2019}. Adversarial learning~\cite{Zhao2018} 
also regularizes the segmentation model by minimizing the distribution difference between segmentation results of annotated images and those of unannotaed images. However, adversarial models are hard to train, and the complex size and shape of PF lesions make it  difficult to capture the true distribution of lesion masks when only a small set of annotated images are available.

\section{Method}
\begin{figure*}[t]
	\centering 
	\includegraphics[width=\textwidth]{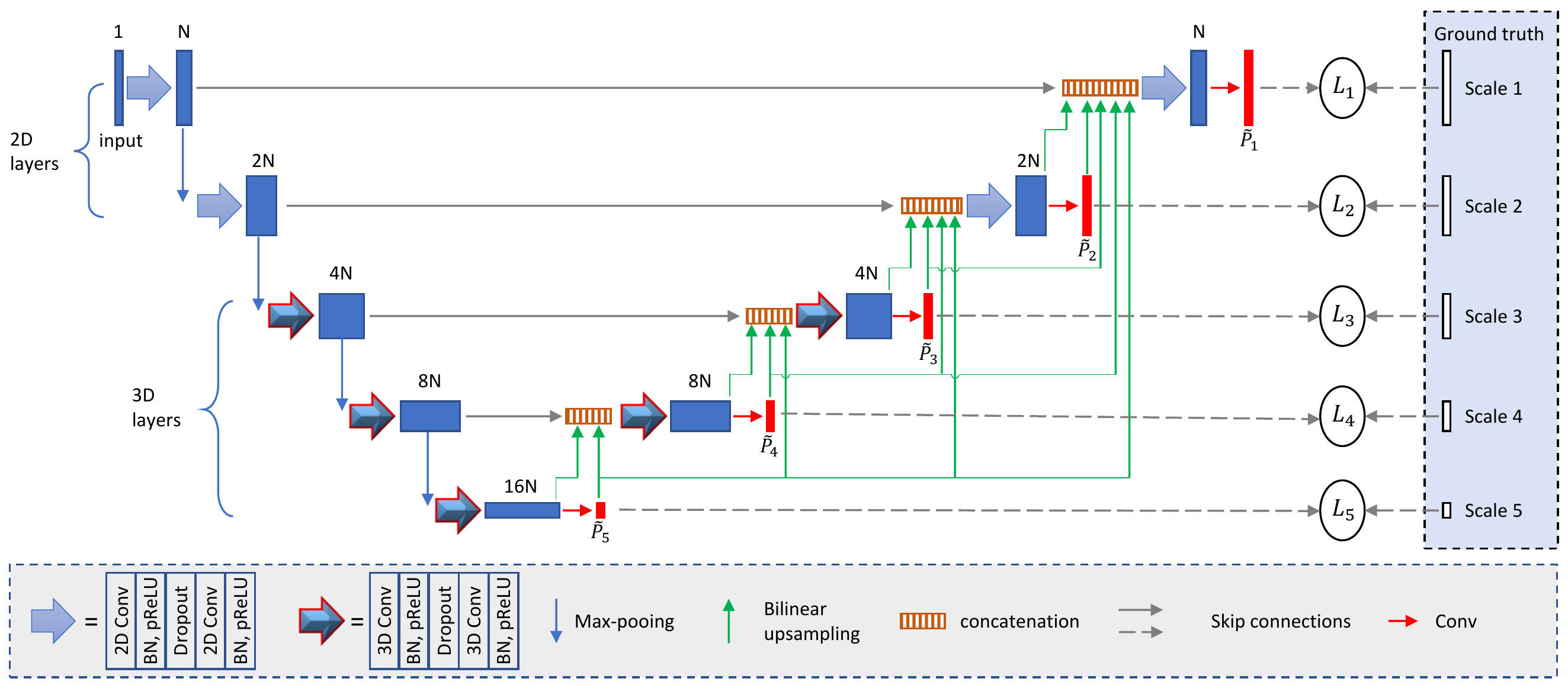}
	\caption{
		Structure of our proposed PF-Net. To deal with 3D images with large inter-slice spacing, the first two scales use 2D convolutions while the other scales use 3D convolutions. $\tilde{P}_s$ is the predicted spatial attention at scale $s$, and it is sent to all lower scales with dense connection in the decoder, as shown by green lines.  }
	\label{fig:network}
\end{figure*}
In this section, we first introduce our proposed Pulmonary Fibrosis segmentation Network (PF-Net) with multi-scale guided dense attention, and then describe how it is used in our I-CRAWL framework for semi-supervised learning.\textbf{}
\subsection{PF-Net: A 2.5D Network with \textcolor{black}{Multi-Scale Guided Dense Attention}}
Our proposed PF-Net is illustrated in Fig.~\ref{fig:network}. It employs an encoder-decoder backbone structure that is commonly used and effective for medical image segmentation~\cite{Ronneberger2015,Milletari2016,Zhou2018}. The encoder contains five scales, and each is implemented by a convolutional block followed by a max-pooling layer for down-sampling. \textcolor{black}{In each block, we use two convolutional layers each followed by a Batch Normalization (BN)~\cite{Ioffe2015} layer} and a parametric Rectified Linear Unit (pReLU), and a dropout layer is inserted before the second convolutional layer. The decoder uses the same type of convolutional blocks as the encoder.  We extend this backbone in the following aspects:
\subsubsection{2.5D Network Structure}
To deal with the anisotropic 3D resolution with high inter-slice spacing and low intra-slice spacing, we combine 2D (i.e., intra-slice) and 3D convolutions so that the network has an approximately balanced physical receptive field along each axis. Let $S$ represent the number of scales in the network ($S=5$ in this paper), we use 2D convolutions and 2D max-poolings for the first $M$ scales in the encoder, and employ 3D convolutions and 3D max-poolings for the last $S-M$ scales in the encoder.  Each resolution level in the decoder contains the same type of 2D or 3D convolutional blocks as in the encoder. We use trilinear interpolation for upsampling in the decoder.

As our lung CT images have a resolution around 0.3$\times$0.3$\times$1.25 mm, i.e., the in-plane resolution is about four times of the through-plane resolution, we set $M=2$ so that the 2D max-pooling layers in the first two scales make the resulted feature maps have a near-isotropic 3D resolution, as shown in Fig.~\ref{fig:network}. Note that the first and $S$-th scales have the highest and lowest spatial resolution, respectively. We use $F^e_s$ and $F^d_s$  to denote the feature maps obtained by the last convolutional block at scale $s$ in the encoder and decoder, respectively.
Note that $F^e_S$ = $F^d_S$ in the bottleneck block.
\subsubsection{\textcolor{black}{Multi-Scale  Guided Dense Attention}}\label{sec:method:mdac}
To better deal with PF lesions with various position and size, we use multi-scale attention to improve the network's spatial awareness, and propose dense attention to leverage multi-scale contextual information for the segmentation task. Specifically, at each scale $s$ of the decoder, we use a convolutional layer to get a spatial attention map $\tilde{P}_s$, and use $\tilde{P}_s$ as a high-level attention signal to guide the learning in all lower scales of the decoder. The input of the decoder at scale $s$ is a concatenation of three parts: $F^e_s$ from scale $s$ of the encoder, an upsampled version of the decoder feature map $F^d_{s+1}$ and $\tilde{P}^s_{s+1}\oplus \tilde{P}^s_{s+2} \oplus ... \oplus \tilde{P}^s_{S}$, where $\oplus$ is the concatenation operation and $\tilde{P}^s_{s+1}$ is the upsampled version of $\tilde{P}_{s+1}$ so that it has the same spatial resolution as $F^e_s$. Therefore, a lower scale  accepts the attention maps from all the higher scales as input, which is referred to as \textcolor{black}{Multi-Scale Dense Attention (MSDA)}. The decoder thus takes advantage of multi-scale contextual information that enables the network to pay more attention to the target region. 

To better learn the spatial attention, we propose \textcolor{black}{Multi-Scale Guided Attention (MSGA)} to explicitly supervise $\tilde{P}_1, \tilde{P}_2, ..., \tilde{P}_S$ at different scales. Let $P_s$ denote the softmaxed output of $\tilde{P}_s$ and $Y$ denote the ground truth (i.e., one-hot probability map) of a training sample $X$. Unlike a common deep supervison strategy that upsamples $\tilde{P_s}$ or $P_s$ at different scales to the same spatial resolution as $Y$~\cite{Dou2016miccai,Zhang2018f,Wang2019a}, we down-sample $Y$ to obtain multi-scale ground truth for spatial attention, which makes the loss calculation more efficient and can directly supervise the spatial attention maps at different scales. Let $Y_s$  denote the down-sampled ground truth at scale $s$. The multi-scale loss function for a single image is:
\begin{equation}\label{eq:loss1}
	\ell(\mathcal{P}, Y) = \sum_{s = 1}^{S}\alpha_s L_s(P_s, Y_s)
\end{equation}
where $\mathcal{P}=\{P_1, P_2, ..., P_S\}$. $L_s()$ is a base loss function for image segmentation, such as the Dice loss~\cite{Milletari2016}.  $\alpha_s$ is the weight of $L_s()$ at scale $s$.

\subsection{Semi-Supervised Learning using I-CRAWL}
\begin{figure*}[t]
	\centering 
	\includegraphics[width=\textwidth]{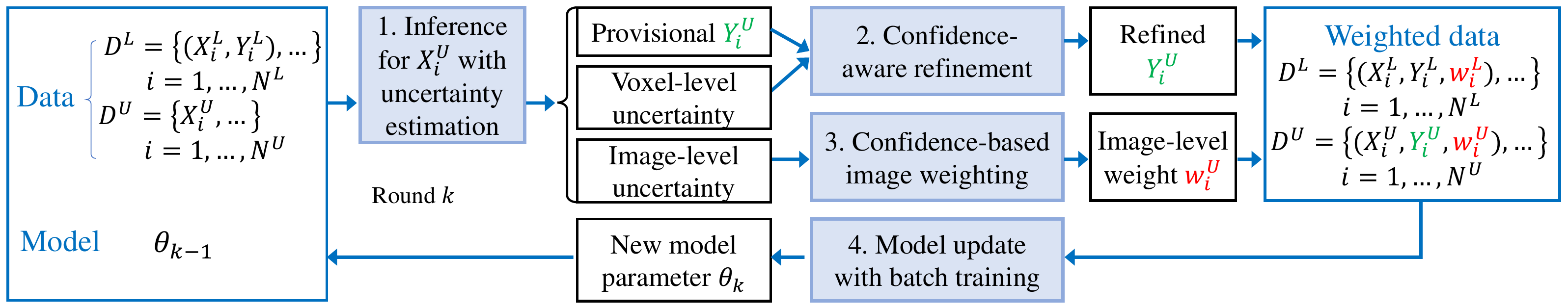}
	\caption{
		Our proposed I-CRAWL (Iterative Confidence-based Refinement And Weighting of pseudo Labels) framework for semi-supervised segmentation. }
	\label{fig:icrawl}
\end{figure*}
Generating pseudo labels for unannotated images has been shown effective for semi-supervised segmentation~\cite{Lee2013,Bai2017a,Fan2020}. However, the pseudo labels often contain some incorrect regions, and low-quality pseudo labels can largely limit the performance of the learned segmentation model. To prevent the learning process from being corrupted by inaccurate pseudo labels, we propose an Iterative Confidence-based Refinement And Weighting of pseudo Labels (I-CRAWL) framework for semi-supervised segmentation. 

Assume that the entire training set consists of one subset $\mathcal{D}^L$ with $N^L$ labeled images and another subset $\mathcal{D}^U$ with $N^U$ unlabeled images. For an image $X^L_i \in \mathcal{D}^L$, its ground truth label $Y^L_i$ is known, while for an image $X^U_i \in \mathcal{D}^U$, its ground truth label is not provided, and we use $Y^U_i$ to denote its estimated pseudo label. As the pseudo labels may have a large range of quality, we introduce an image-level weight $w^U_i \in [0, 1]$ for each pair of $X^U_i,Y^U_i$ for learning, and define $w^U_i$ based on the confidence (or uncertainty, inversely) of $Y^U_i$. The weight for a labeled image $X^L_i$ from $\mathcal{D}^L$ can be similarly denoted as $w^L_i$, and we set $w^L_i = 1$ as the corresponding label $Y^L_i$ is clean and reliable. Therefore, the labeled subset can be denoted as $\mathcal{D}^L = \{(X^L_i, Y^L_i, w^L_i), ...\}$ ($i=1, 2, ..., N^L$), and the unlabeled subset with pseudo labels can be denoted as $\mathcal{D}^U = \{(X^U_i, Y^U_i, w^U_i), ...\}$ ($i=1, 2, ..., N^U$).

Our I-CRAWL is illustrated in Fig.~\ref{fig:icrawl}, and it is an iterative learning process with $K$ rounds. Each round has four steps: 1) inference for unannotated images with uncertainty estimation, 2) confidence-aware refinement of pseudo labels, 3) confidence-based image weighting, and 4) network update where the current pseudo labels and image-level weights are used to train the network. These steps are detailed as follows.
\subsubsection{Inference for Unannotated Images with Uncertainty Estimation}
 With the network parameter $\theta_{k-1}$ obtained in the last round, in round $k$, we first use $\theta_{k-1}$ to predict provisional pseudo labels for unannotated images in $\mathcal{D}^U$ and the associated uncertainty estimation.  
 Note that in the first round (i.e., $k = 1$), the initial network parameter  $\theta_0$ is obtained by pre-training with the annotated images in $\mathcal{D}^L$.
 
 With $\theta_{k-1}$, we employ the Monte Carlo (MC) Dropout~\cite{Gal2016} that has shown to be an effective method for estimation of epistemic uncertainty caused by the lack of training data~\cite{Yu2019,Xia2020}. MC Dropout feeds an image $X^U_i$ into the network $R$ times with random dropout, which leads to $R$ predictions, i.e., $R$ foreground probability maps~\cite{Gal2016}. The average of these $R$ foreground probability maps is taken as the provisional probability map  $P^U_i$, a binarization of which leads to a provisional pseudo segmentation label $Y^U_i$. 

At the same time, the statistical variance of the $R$ foreground probability maps is taken as the uncertainty map $V^U_i$, which gives voxel-level uncertainty. As uncertainty information can indicate  potentially wrong segmentation results~\cite{Yu2019,Wang2019i}, we treat the pseudo labels with high uncertainty (i.e., low confidence) values as unreliable labels for images in $\mathcal{D}^U$, and propose a Confidence-Aware  Refinement (CAR) method to improve the pseudo labels' quality. 

\subsubsection{Confidence-Aware Refinement of Pseudo Labels} 
Given the provisional pseudo label $Y^U_i$ with the uncertainty map $V^U_i$ for an unannotated image $X^U_i\in\mathcal{D}^U$, and let $\mathbf{x}$ denote a voxel, we split the voxels in $Y^U_i$ into three sets according to the status: high-confidence foreground voxels $\mathbb{F} = \{\mathbf{x}| Y^U_{i\mathbf{x}} = 1, V^U_{i\mathbf{x}} \le \epsilon\}$, high-confidence background voxels $\mathbb{B} = \{\mathbf{x}| Y^U_{i\mathbf{x}} = 0, V^U_{i\mathbf{x}} \le \epsilon\}$ and undetermined voxels $\mathbb{U} = \{\mathbf{x}|V^U_{i\mathbf{x}} > \epsilon\}$, where $\epsilon=0.05$ is a small threshold value.
For undetermined voxels in $\mathbb{U}$, we refine their labels according to a contextual regularization considering inter-voxel connections and softened probabilities for these voxels. 

Our CAR for pseudo label refinement  has two steps:  probability map softening and contextual regularization. First, we soften the foreground probabilities for uncertain voxels, which will degrade the influence of the network's prediction for these voxels in the following contextual regularization step. Let $p$ denote a foreground probability value and $u$ denote the corresponding uncertainty value, the softening function is:
\begin{equation}
	g(p,u)=0.5 + (p - 0.5)\times(1-u)
\end{equation}
where the softened foreground probability get closer to 0.5 when $u$ is larger. Let $\dot{P}^U_{i\mathbf{x}}$ denote the softened foreground probability of voxel $\mathbf{x}$, and is obtained by:
\begin{align}
	\dot{P}^U_{i\mathbf{x}}=\begin{cases}
	P^U_{i\mathbf{x}} & \text{if } {\mathbf{x}\in \mathbb{F}\cup\mathbb{B}}
		\\
	g(P^U_{i\mathbf{x}},V^U_{i\mathbf{x}}) & \text{if } {\mathbf{x}\in \mathbb{U}}
	\end{cases}
\end{align}
 
Then, we use contextual regularization taking $\dot{P}^U_i$ as input to refine the pseudo labels, which is implemented by a fully connected Conditional Random Field (CRF)~\cite{Krahenbuhl2011}. For simplicity, we denote $X^U_{i\mathbf{x}}$, $Y^U_{i\mathbf{x}}$ and  $\dot{P}^U_{i\mathbf{x}}$ as $x_{\mathbf{x}}$, $y_{\mathbf{x}}$ and $\dot{p}_{\mathbf{x}}$, respectively. The energy function of CRF is:
\begin{align}\label{eq:crf}
	E(Y^U_i)=\sum_{\mathbf{x}}\phi(y_{\mathbf{x}}) + 
	\sum_{\mathbf{x,y}}\psi(y_{\mathbf{x}},y_{\mathbf{y}})
\end{align}
where  $\phi(y_{\mathbf{x}}) = -y_{\mathbf{x}}\text{log}(\dot{p}_{\mathbf{x}}) - (1-y_{\mathbf{x}})(1-\dot{p}_{\mathbf{x}})$ constrains the output to be consistent with the softened probability map, and this constraint for the low-confidence voxels in $\mathbb{U}$ is weak. The second term in Eq.~\eqref{eq:crf} is a pairwise potential that encourages the label's contextual consistency:
\begin{multline}\label{eq:crf_psi}
	\psi(y_{\mathbf{x}},y_{\mathbf{y}}) = 
	\mu(y_{\mathbf{x}},y_{\mathbf{y}})
	 \Big[ w_1\text{exp}(-\frac{||\mathbf{x} - \mathbf{y}||^2}{2\sigma_\alpha^2}-\frac{||x_{\mathbf{x}} - x_{\mathbf{y}}||^2}{2\sigma_\beta^2}) \\
	+w_2\text{exp}(-\frac{||\mathbf{x} - \mathbf{y}||^2}{2\sigma_\gamma^2}) \Big]
\end{multline}
where  $\mu(y_{\mathbf{x}},y_{\mathbf{y}})=1$ if $y_{\mathbf{x}}\neq y_{\mathbf{y}}$ and 0 otherwise. Minimization of Eq.~\eqref{eq:crf} leads to a refined pseudo label $Y^U_i$ for $X^U_i$.

\subsubsection{Confidence-based \textcolor{black}{Image Weighting} of Pseudo Labels}\label{sec:method:weight}
With the new pseudo label $Y^U_i$, we further employ the confidence to update its \textcolor{black}{image-level} weight $w^U_i$ to suppress low-quality pseudo labels at the image level. We first define an \textcolor{black}{image-level uncertainty} $\mathbf{v}_i$ based on uncertainty map  $V^U_i$:
\begin{align}
	\mathbf{v}_i = \frac{\sum_{\mathbf{x}} V^U_{i\mathbf{x}} }{\sum_{\mathbf{x}} Y^U_{i\mathbf{x}} + \eta}
\end{align}
where $\mathbf{v}_i$ is the sum of voxel-level uncertainty normalized by the segmented lesion's volume in the image. $\eta = 10^{-5}$ is a small number for numerical stability. Let $v_{max}$ and $v_{min}$ denote the maximal and minimal values of $\mathbf{v}_i$ among all the unannotated images, we  map $\mathbf{v}_i$ to the range of [0, 1]:
\begin{align}
	\mathbf{v}'_i = \frac{\mathbf{v}_i - \mathbf{v}_{min} }{\mathbf{v}_{max} - \mathbf{v}_{min} }
\end{align}

Finally, the \textcolor{black}{image-level weight} $w^U_i$ is defined as:
\begin{align}\label{eq:weight}
	w^U_i = (1.0 - \mathbf{v}'_i)^{1/\gamma}
\end{align}
where $\gamma \geq 1.0$ is a hyper parameter to control the non-linear mapping between the image-level uncertainty $\mathbf{v}'_i$ and the weight. We do not use $\gamma < 1.0$ as it will lead the weight of most samples to be very small (close to 0.0).
In contrast, $\gamma > 1.0$ leads the weight for most samples close to 1.0, and only samples with a high uncertainty will be strongly suppressed. In the experiment, we set $\gamma = 3.0$ according to the best performance on the validation set. 

\subsubsection{Model Update with Batch Training}
With the refined pseudo labels and image-level weights obtained above, we train the network based on $\mathcal{D}^L$ and the current pseudo labels for images in $\mathcal{D}^U$, where each image is weighted by $w^{L}_i$ or $w^{U}_i$ in the segmentation loss function. The weighted loss for the entire training set is:
\begin{equation}\label{eq:loss2}
	\mathcal{L}= \sum_{i = 1}^{N^L}w^L_i \ell(\mathcal{P}^L_i, Y^L_i) + 
	\sum_{i = 1}^{N^U}w^U_i \ell(\mathcal{P}^U_i, Y^U_i)
\end{equation}
where $\ell()$ is defined in Eq.~\eqref{eq:loss1}. $\mathcal{P}^L_i$ and $\mathcal{P}^U_i$ are the multi-scale predictions obtained by PF-Net for an annotated image and  an unannotated image, respectively.


\section{Experiments and Results}
\subsection{Experimental Setting}
\subsubsection{Data}
Thoracic CT scans of 41 Male Rhesus Macaques with radiation-induced lung damage were collected with ethical committee  approval. Once irradiated, each individual underwent serial CT scans for assessment of PF  around every 30 days in 3 to 8 months. 133 CT scans with PF were used for experiments of the segmentation task. The CT scans have a slice thickness of 1.25~mm, with image size 512 $\times$ 512 and pixel size ranging from 0.20~mm $\times$ 0.20~mm  to 0.38~mm $\times$ 0.38~mm. We randomly split the dataset at individual level into 86 scans from 27 individuals for training, 15 scans from 4 individuals for validation, and 32 scans from 10 individuals for testing. Manual annotations given by an experienced radiologist were used as the segmentation ground truth. In the training set, we used 18 scans with annotations as $\mathcal{D}^L$ and the other 68 scans as unannotated images $\mathcal{D}^U$ for semi-supervised learning, and also investigated the performance of our method with different ratios of annotated images. For preprocessing, we crop the lung region and normalize the intensity to [0, 1] using a window/level of 1500/-650.

\subsubsection{Implementation and Evaluation Metrics}
Our PF-Net\footnote{\url{https://github.com/HiLab-git/PF-Net}} and I-CRAWL framework were implemented in Pytorch with PyMIC\footnote{\url{https://github.com/HiLab-git/PyMIC}}~\cite{Wang2020c} library on a Ubuntu desktop with an NVIDIA GTX 1080 Ti GPU. The channel number parameter $N$ in PF-Net was set as 16. Dropout was only used in the encoder. The dropout rate at the first two scales of the encoder was  0 due to the small number of feature channels, and that for the last three scales was 0.3, 0.4 and 0.5, respectively. 
We set the base loss function $L_s()$ as Dice loss~\cite{Milletari2016}. PF-Net was trained with Adam optimizer, weight decay of $10^{-5}$, patch size of $48 \times 192 \times 192$ and batch size of 2. 

 \textcolor{black}{The round number for our I-CRAWL was $K = 3$, and we use round 0 to refer to the pre-training with annotated images. In each of the following round, the pseudo labels of unannotated images acquired by CAR were kept fixed, and we used Adam optimizer to train the network for tens of thousands of iterations until the performance on validation set stopped to increase.} The learning rate was initialized as $10^{-3}$ and halved every 10k iterations. Uncertainty estimation was based on 10 forward passes of MC dropout. We used the SimpleCRF library\footnote{\url{https://github.com/HiLab-git/SimpleCRF}} to implement fully connected CRF~\cite{Krahenbuhl2011}. Following~\cite{Kamnitsas2017}, image intensity was rescaled from [0, 1] to [0, 255] before the image was sent into the CRF, and the CRF parameters were: $w_1=3$, $w_2=10$, $\sigma_\alpha=10$, $\sigma_\beta=20$ and $\sigma_\gamma=15$, which was tuned based on the validation set. $\gamma$ in Eq.~\eqref{eq:weight} was 3.0, and the performance based on different  $\gamma$ values is shown in Table~\ref{tab:valid_gamma}.

For quantitative evaluation of the segmentation, we used Dice score, Relative Volume Error (RVE) and the 95-th percentile of Hausdorff Distance (HD$_{95}$) between the segmented PF lesions and the ground truth in 3D volumes. Paired t-test was used to see if two methods were significantly different.

\subsection{Performance of PF-Net}
In this section, we investigate the performance of our PF-Net for pulmonary fibrosis segmentation only using the 18 annotated images for training. The results of semi-supervised learning will be demonstrated in Section~\ref{sec:result_semi}. 

\subsubsection{Comparison with Existing Networks}
\begin{figure*}[t]
	\centering 
	\includegraphics[width=\textwidth]{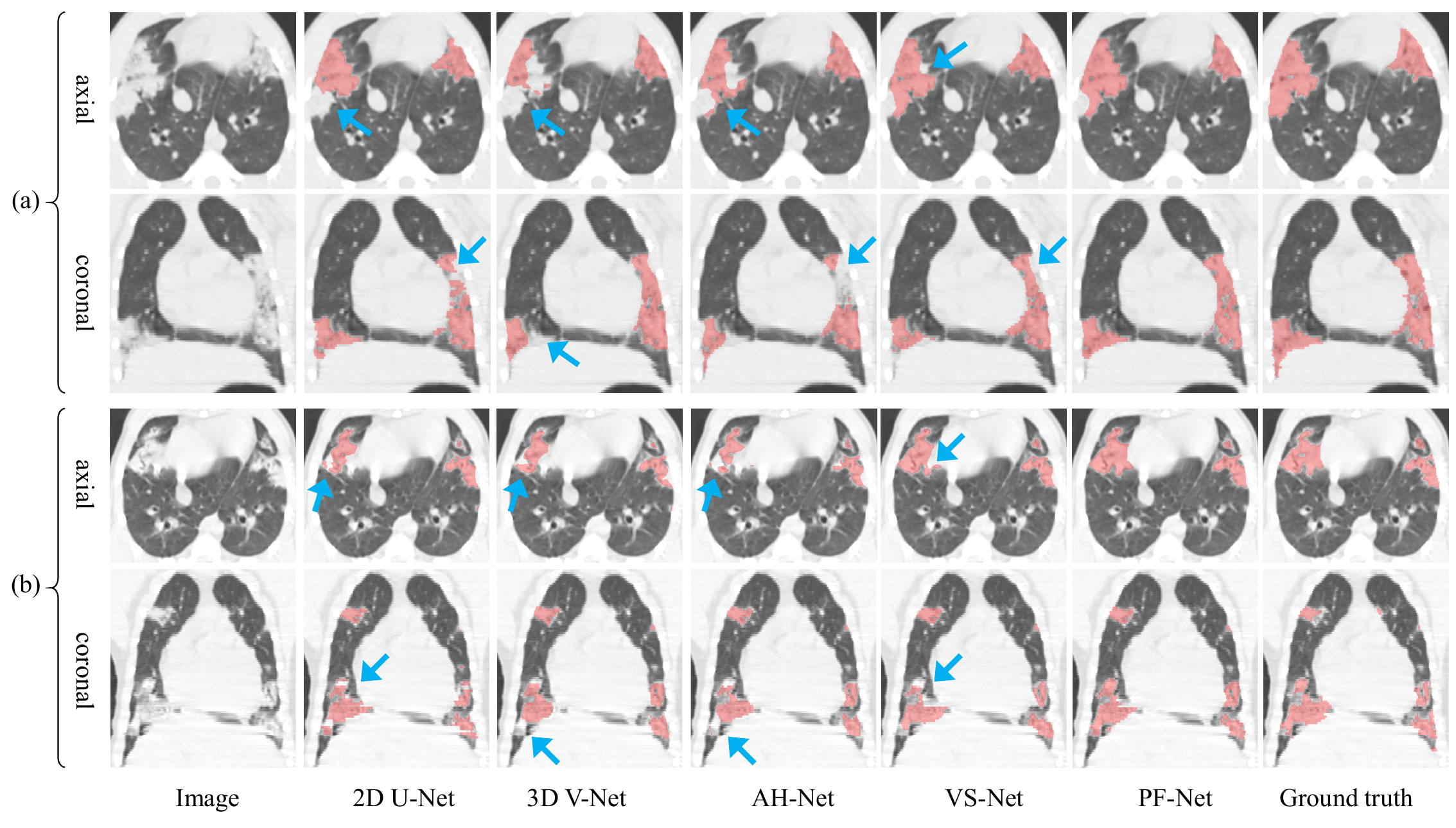}
	\caption{
		Qualitative comparison between different networks for PF lesion segmentation.  (a) and (b) are from two different individuals. Blue arrows indicate some mis-segmented regions.}
	\label{fig:net_compare}
\end{figure*}
\begin{table}
	\centering
	\caption{Quantitative comparison of different networks for PF lesion segmentation. \textcolor{black}{$^*$ denotes significant difference from the others ($p$-values $<$ 0.05).}
	}
	\label{tab:net_compare_others}
	\begin{tabular}
		{l|c|c|c} 
		\hline
		& Dice (\%) & RVE (\%)  & HD$_{95}$ (mm) \\ \hline
		2D U-Net~\cite{Ronneberger2015} & 59.31$\pm$15.66 & 38.05$\pm$19.17 & 15.73$\pm$15.57 
		\\  
		Attention U-Net~\cite{Oktay2018} &  59.22$\pm$14.55 &31.93$\pm$20.47 & 18.64$\pm$15.58
		\\   \hline
		3D U-Net~\cite{Abdulkadir2016}  & 57.19$\pm$13.47 & 40.09$\pm$22.45 &  18.93$\pm$12.98 
		\\ 
		3D scSE-Net~\cite{Roy2019} &  59.54$\pm$12.24 & 41.95$\pm$19.59 &  19.18$\pm$15.57
		\\ 
		3D V-Net~\cite{Milletari2016}  &  60.37$\pm$13.33 & 35.03$\pm$27.95 &   19.22$\pm$14.65	
		\\ \hline
		\textcolor{black}{nnU-Net~\cite{Isensee2021}} &  64.51$\pm$15.25  & 33.19$\pm$21.54  &  18.46$\pm$15.70 \\ 
		AH-Net~\cite{Liu2018c} &  60.50$\pm$13.27  & 35.77$\pm$23.96  &  17.90$\pm$13.66 \\
		VS-Net~\cite{Wang2019b} & 67.45$\pm$9.70 & 30.03$\pm$24.78 &  18.35$\pm$17.67  
		\\ 
		PF-Net & \textbf{70.36$\pm$10.14$^*$} &  \textbf{27.96$\pm$21.72}  & \textbf{10.87$\pm$9.28$^*$}  \\
		\hline
	\end{tabular}
\end{table}
Our PF-Net was compared with three different categories of network structures: 1) 2D CNNs including the typical 2D U-Net~\cite{Ronneberger2015} and more advanced attention U-Net~\cite{Oktay2018} that leverages spatial attention to focus more on the segmentation target; 2) 3D CNNs including 3D U-Net~\cite{Abdulkadir2016}, 3D V-Net~\cite{Milletari2016} and scSE-Net~\cite{Roy2019} that combines a 3D U-Net~\cite{Abdulkadir2016} backbone with spatial and channel ``squeeze and excitation" modules; and 3) existing networks designed for dealing with volumetric images with large inter-slice spacing (i.e., anisotropic resolution): \textcolor{black}{nnU-Net~\cite{Isensee2021} that automatically configures the network structure so that it is adapted to the given dataset,} AH-Net~\cite{Liu2018c} that transfers features learned from 2D images to 3D anisotropic volumes, and VS-Net~\cite{Wang2019b} that uses a mixture of 2D and 3D convolutions with spatial attention. Quantitative evaluation results of these methods are shown in Table~\ref{tab:net_compare_others}. We found that 3D U-Net~\cite{Abdulkadir2016} achieved a Dice score of 57.19\%, which was the lowest among the compared methods. 2D U-Net~\cite{Ronneberger2015}, Attention U-Net~\cite{Oktay2018}, 3D V-Net~\cite{Milletari2016} and 3D scSE-Net~\cite{Roy2019} achieved similar performance with Dice score around 60\%. The nnU-Net~\cite{Isensee2021}, AH-Net~\cite{Liu2018c} and VS-Net~\cite{Wang2019b} designed to deal with anisotropic resolution generally performed better than these 2D and 3D networks. Among these existing methods, VS-Net achieved the best Dice that was 67.45\%. Our PF-Net  achieved Dice, RVE and HD$_{95}$ of 70.36\%, 27.96\% and 10.87 mm, respectively, where the Dice and HD$_{95}$ were significantly better than those of the other compared methods.

Fig.~\ref{fig:net_compare} shows a qualitative comparison between these networks, where (a) and (b) are from two individuals, and axial  and coronal views are shown in each case. In Fig.~\ref{fig:net_compare}(a), it can be seen that 2D U-Net~\cite{Ronneberger2015}, 3D V-Net~\cite{Milletari2016} and AH-Net~\cite{Liu2018c} lead to obvious  under-segmentation, as highlighted by blue arrows in the first row. VS-Net~\cite{Wang2019b} performs better than them, but the result of our PF-Net is closer to the ground truth than that of VS-Net. From the coronal view of Fig.~\ref{fig:net_compare}(b), we can observe that the result of 2D U-Net~\cite{Ronneberger2015} lacks inter-slice consistency. 3D V-Net~\cite{Milletari2016} achieves better inter-slice consistency, but it has a poor segmentation in the upper and lower regions of the lesion, as indicated by the blue arrow in the third column. AH-Net~\cite{Liu2018c}, VS-Net~\cite{Wang2019b} and our PF-Net that consider anisotropic resolution obtain better performance than the above networks purely using 2D or 3D convolutions. What's more, Fig.~\ref{fig:net_compare} shows that the lesions have complex and irregular sizes and shapes, and  the proposed PF-Net is able to achieve more accurate segmentation in these cases than AH-Net~\cite{Liu2018c} and VS-Net~\cite{Wang2019b}.

\subsubsection{Ablation Study}
\begin{figure*}[t]
	\centering 
	\includegraphics[width=\textwidth]{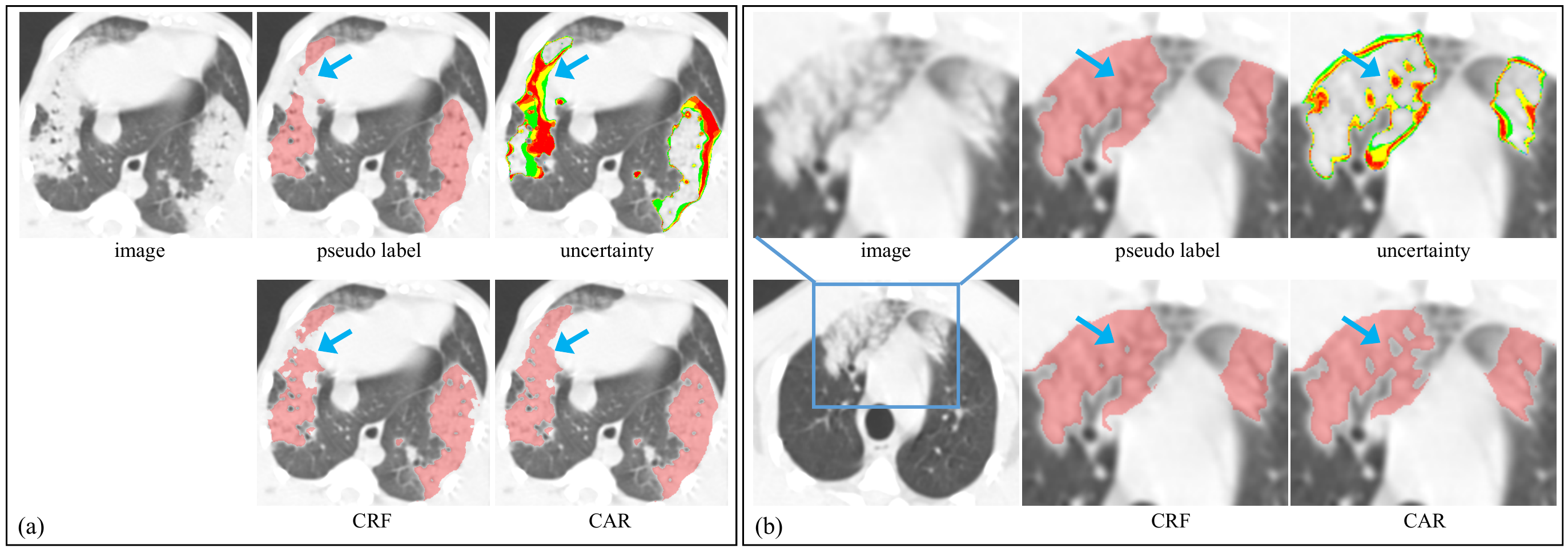}
	\caption{
		Visual comparison between CRF and our proposed Confidence-Aware Refinement (CAR) for update of pseudo labels. In the uncertainty map, dark green and red colors represent low and high uncertainty values, respectively. Blue arrows highlight some local differences. }
	\label{fig:crf_compare}
\end{figure*}
\begin{table}
	\centering
	\caption{Quantitative comparison of different variants of baseline network and our PF-Net. \textcolor{black}{Baseline ($M$) means 2D and 3D convolutions are used in the first $M$ and the other convolutional blocks, respectively. The second section is based on Baseline ($M$ = 2), and $^*$ denotes significant improvement from it ($p$-value $<$ 0.05).} Note that the last row is our PF-Net. 
	}
	\label{tab:net_compare_ablation}
	\begin{tabular}
		{l|c|c|c} 
		\hline
		& Dice (\%) & RVE (\%)  & HD$_{95}$ (mm) \\ \hline
			\textcolor{black}{Baseline ($M$ = 0)} & 60.83$\pm$14.93 & 36.09$\pm$21.29 & 19.80$\pm$11.48 
		\\ 
			\textcolor{black}{Baseline ($M$ = 1)} & 62.42$\pm$13.01 & 33.74$\pm$24.50 & 18.87$\pm$12.03 
		\\ 
			Baseline ($M$ = 2) & \textbf{67.87$\pm$9.96} & \textbf{30.79$\pm$20.37} & \textbf{15.31$\pm$15.88} 
		\\ 
			\textcolor{black}{Baseline ($M$ = 3)} & 65.98$\pm$12.27 & 31.45$\pm$21.84 & 16.92$\pm$12.66 
		\\ 
			\textcolor{black}{Baseline ($M$ = 4)} & 65.93$\pm$14.01 & 32.37$\pm$23.17 & 17.35$\pm$12.67 
		\\ 
		\textcolor{black}{Baseline ($M$ = 5)} & 63.11$\pm$10.90 & 32.71$\pm$22.35 & 17.60$\pm$12.28 
		\\  \hline
		\textcolor{black}{+ non-local} & 68.63$\pm$15.49 & 29.82$\pm$22.27 &  14.80$\pm$13.99
		\\
		+ deep supervision & 68.57$\pm$10.30 & 30.73$\pm$21.24 &  17.56$\pm$19.22
		\\ 
		+ MSGA & 69.32$\pm$13.20 & \textbf{27.55$\pm$21.67} &  13.85$\pm$16.68  
		\\
		+ MSGA + MSDA$^\circ$ & 70.25$\pm$10.09$^*$ & 28.09$\pm$19.67  &  12.28$\pm$12.12 \\
		+ MSGA + MSDA & \textbf{70.36$\pm$10.14$^*$} & 27.96$\pm$21.72   &  \textbf{10.87$\pm$9.28$^*$}\\
		\hline 
	\end{tabular}
\end{table}
To investigate the effectiveness of each component of our PF-Net, we set the baseline as a naive 2.5D U-Net that extends 2D U-Net~\cite{Ronneberger2015} by using pReLU, replacing 2D convolutions with 3D convolutions at the three lowest resolution levels, and adding dropout layers to each convolutional block. \textcolor{black}{To justify the choice of using 2D convolutions at the first two resolution levels and 3D convolutions at the other resolution levels, we set $M$ to 0-5 respectively. Note that $M = 0$ and $M = 5$ correspond to pure 2D and pure 3D networks, respectively. Comparison between these variants are listed in the first section of Table~\ref{tab:net_compare_ablation}, which shows that the performance increases as $M$ changes from 0 to 2, and decreases when $M$ is 3 and larger. This is in line with our motivation to set $M$ to 2 due to the fact that the in-plane resolution is around four times of the through-plane resolution. Thus, we use baseline ($M$ = 2) in the following ablation study.} 

The proposed PF-Net is referred to as baseline + \textcolor{black}{MSDA + MSGA, where MSDA is our proposed multi-scale dense attention and MSGA is the proposed multi-scale guided attention.} We compared the baseline and our PF-Net with: 1) \textcolor{black}{Baseline + non-local~\cite{Wang2017c}, where the non-local is a self-attention block inserted at the bottleneck (scale 5) of the baseline network, and it was not used at the lower scales with higher resolution due to memory constraint; } 2) Baseline + deep supervision, where the baseline network was only combined with a typical deep supervision strategy as implemented in~\cite{Dou2016miccai,Zhang2018f}; \textcolor{black}{3) Baseline + MSGA, without using MSDA; 4) Baseline + MSGA + MSDA$^\circ$, where  MSDA$^\circ$ is a variant of MSDA} and it refers to  $\tilde{P}_{s}$ is only sent to its next lower scale $s-1$, rather than all the lower scales in the decoder of PF-Net.

\begin{figure}[t]
	\centering 
	\includegraphics[width=0.48\textwidth]{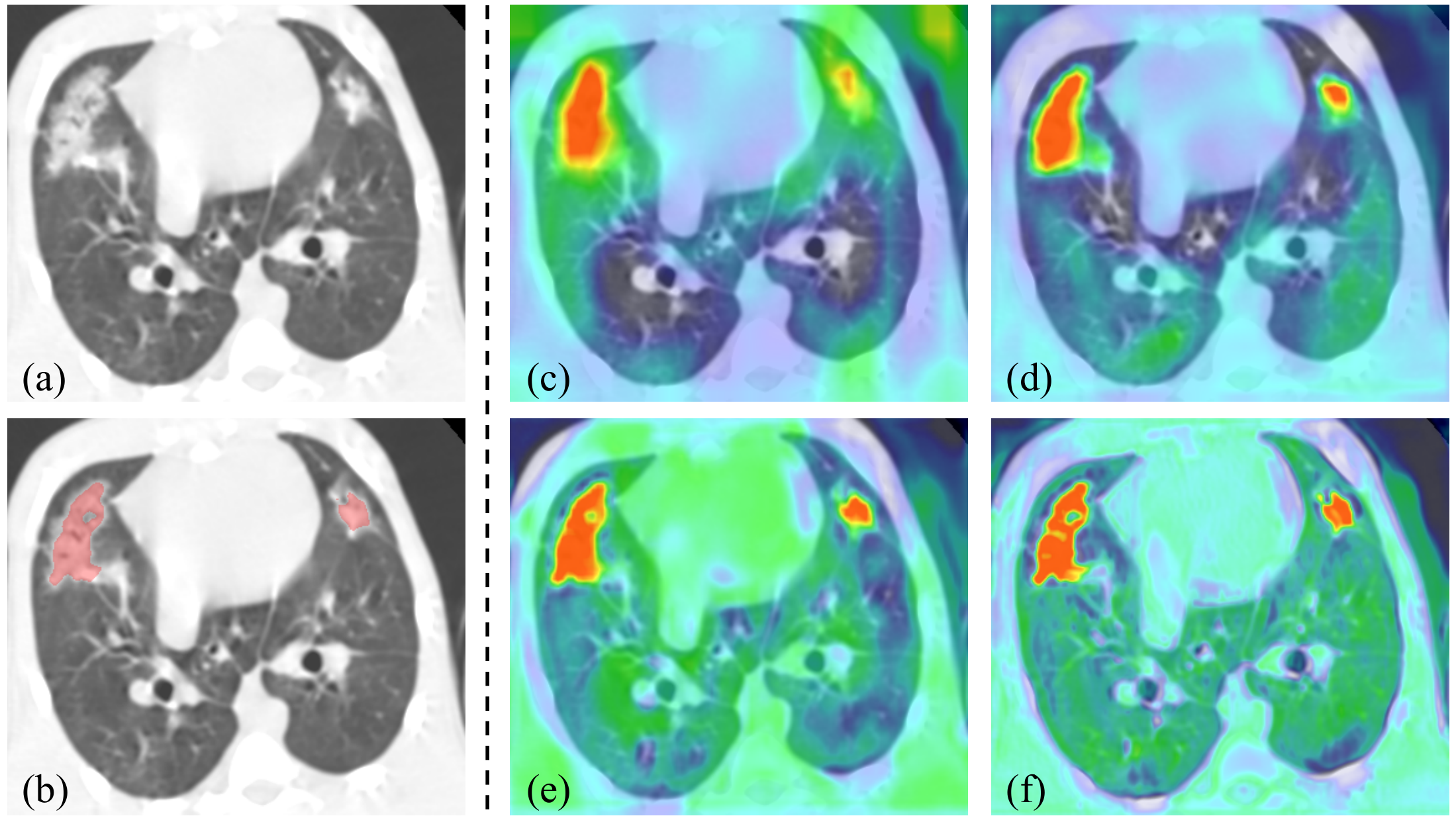}
	\caption{
		\textcolor{black}{Visualization of multi-scale attention maps of PF-Net. (a)-(b) CT image and segmentation result. (c)-(f) Attention maps $~\tilde{P}_5$,  $~\tilde{P}_4$,  $~\tilde{P}_3$ and  $~\tilde{P}_2$, respectively.}}
	\label{fig:attention}
\end{figure}

Quantitative evaluation results of these variants are listed in the second section of Table~\ref{tab:net_compare_ablation}. It shows that 
compared with the baseline ($M = 2$), deep supervision improved the Dice score from 67.87\% to 68.57\%, but our MSGA was more effective with Dice of 69.32\%. Using MSDA$^\circ$ could further improve the performance, but it was less effective than our MSDA. Table~\ref{tab:net_compare_ablation} demonstrates that our PF-Net was better than the compared variants, and it significantly outperformed the baseline in terms of Dice and HD$_{95}$.  \textcolor{black}{Fig.~\ref{fig:attention} presents a visualization of multi-scale attention maps of PF-Net. It shows that attention maps across scales are consistent with each other and they change from coarse to fine as the spatial resolution increases. }

\begin{figure}
	\centering 
	\includegraphics[width=0.48\textwidth]{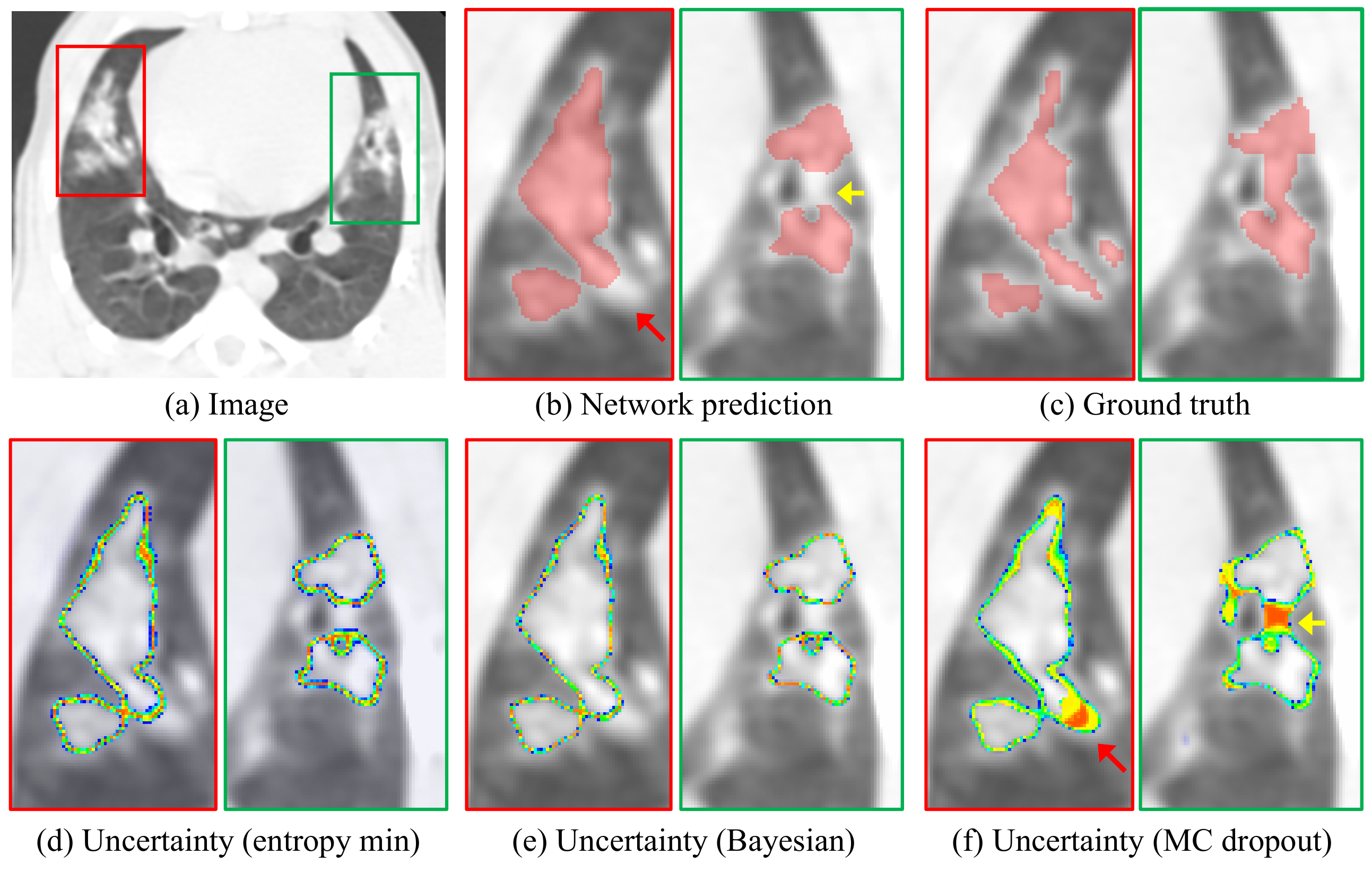}
	\caption{
		\textcolor{black}{Visual comparison of uncertainty obtained by different methods. (b)-(f) show two sub-regions from (a), as indicated by red and green rectangles, respectively.}}
	\label{fig:uncertain_compare}
\end{figure}

\begin{figure*}
	\centering 
	\includegraphics[width=0.8\textwidth]{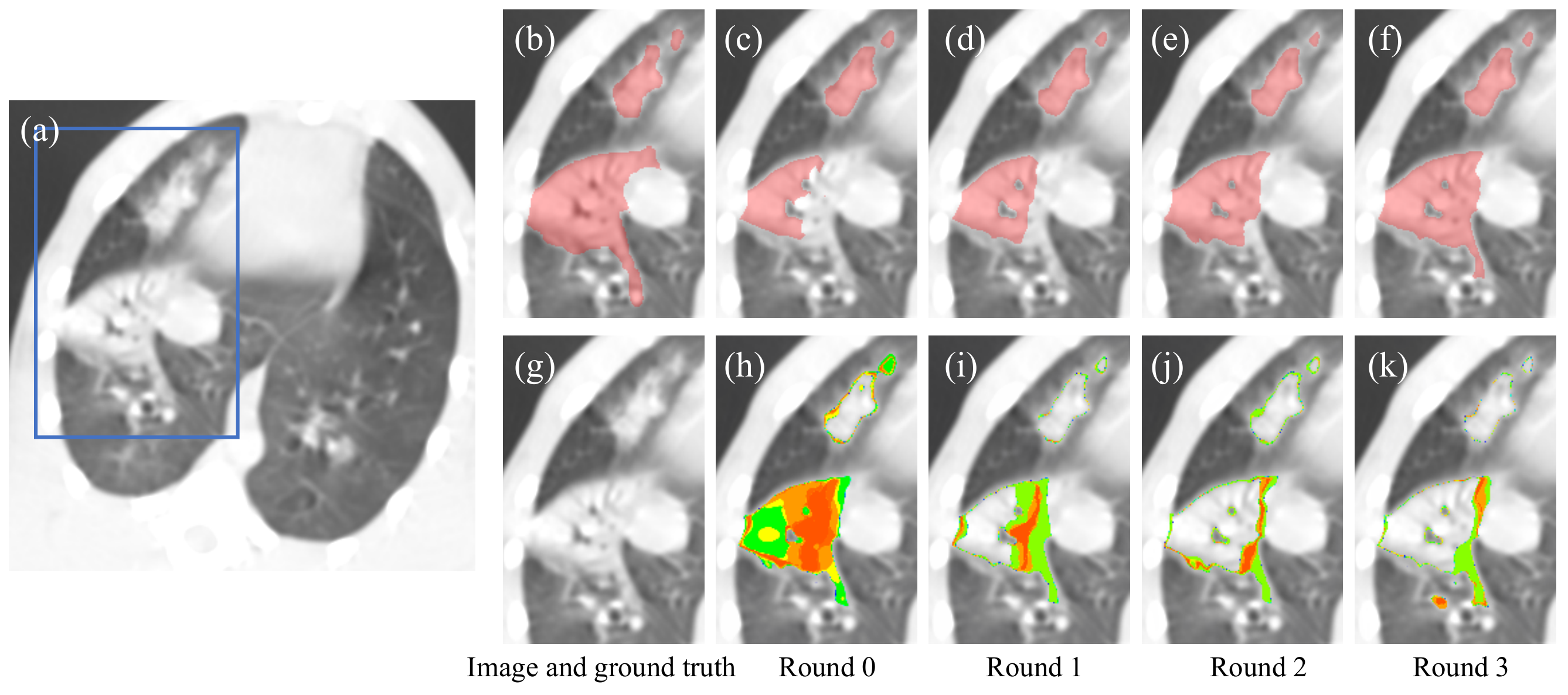}
	\caption{
		\textcolor{black}{Change of uncertainty maps at different rounds of I-CRAWL. (a) input image. (g) sub-region of the input. (b) ground truth (just for reference here, not used during training). (c)-(f) and (h)-(k) show the pseudo labels and the corresponding uncertainty maps of I-CRAWL as the round increases. }}
	\label{fig:uncertain_rounds}
\end{figure*}

\subsection{Results of Semi-supervised Learning}\label{sec:result_semi}
\begin{table}
	\centering
	\caption{Performance on the validation set based on different $\gamma$ values in the confidence-based sampling weighting of pseudo labels in round 1 of I-CRAWL. ``Initial" refers to model pre-trained on annotated images (round 0). \textcolor{black}{$^*$ denotes significant improvement from it ($p$-value $<$ 0.05).}}
	\label{tab:valid_gamma}
	\begin{tabular}
		{l|c|c|c} 
		\hline
		& Dice (\%) & RVE (\%)  & HD$_{95}$ (mm) \\ \hline
		Initial & 66.62$\pm$8.53 & 23.91$\pm$15.14 &  11.11$\pm$6.00
		\\ 
		No weighting & 67.83$\pm$8.25 & 22.74$\pm$17.82 & 9.12$\pm$4.48
		\\ 
		$\gamma = 1.0$ & 67.36$\pm$8.75 & 24.73$\pm$18.83 &  10.02$\pm$5.24  
		\\ 
		$\gamma = 2.0$ & 68.00$\pm$8.23 & 23.01$\pm$16.90  &  \textbf{8.69$\pm$4.60$^*$} \\
		$\gamma = 3.0$ & \textbf{68.75$\pm$8.00$^*$} & \textbf{22.42$\pm$17.11}   &  8.97$\pm$4.45$^*$\\
		$\gamma = 4.0$ & 68.54$\pm$8.64$^*$ & 22.45$\pm$18.06   &  8.87$\pm$4.72$^*$\\
		$\gamma = 5.0$ & 66.86$\pm$8.74 & 23.44$\pm$17.95   &  8.75$\pm$4.51$^*$\\
		\hline
	\end{tabular}
\end{table}

With PF-Net as the segmentation network structure, we further validate our proposed I-CRAWL for semi-supervised training. In Section~\ref{sec:icrawl_hyper-para} and \ref{sec:result_ssl_car} we used 18 annotated and 68 unannounced images, i.e., the annotation ratio was around 20\%. In Section~\ref{sec:result_ssl_ablation}, we  experimented with different annotation ratios including 10\%, 20\% and 50\%.
\subsubsection{Hyper-Parameter Setting}\label{sec:icrawl_hyper-para}
First, to investigate the best value of hyper parameter $\gamma$ in Eq.~\eqref{eq:weight} that controls the weight of unannotated images, we measured the model's performance on the validation set at the first round of I-CRAWL with $\gamma$ ranging from 1.0 to 5.0. They are compared with ``Initial" that refers to model pre-trained on the annotated images and ``No weighting" that denotes treating all unannotated images equally without considering the quality of pseudo labels. Quantitative measurements shown in Table~\ref{tab:valid_gamma} demonstrate that the best $\gamma$ value was 3.0, and its corresponding Dice score was 68.75\%, which was better than 66.62\% obtained by ``Initial" and 67.83\% obtained by ``No weighting". Therefore, we set $\gamma=3.0$ in the following experiments.
\subsubsection{Uncertainty and Confidence-Aware Refinement}\label{sec:result_ssl_car}
Fig.~\ref{fig:crf_compare} shows a visual comparison  between a standard  fully connected CRF~\cite{Krahenbuhl2011} and our proposed CAR that leverages confidence (uncertainty) for refinement of pseudo labels. The first row of each subfigure shows an unannotated image and the pseudo label with uncertainty obtained by the CNN, and the second row shows the updated pseudo labels. In Fig.~\ref{fig:crf_compare}(a), the initial pseudo label has a large under-segmentated region, which is associated with high values in the uncertainty map, i.e., low confidence. The standard CRF only fixed the pseudo label moderately. In contrast, with the help of confidence, our CAR largely improved the pseudo label's quality by recovering the under-segmented region. In Fig.~\ref{fig:crf_compare}(b), the initial pseudo label has some over-segmentation in airways, and the uncertainty map indicates the potentially wrong segmentation in the corresponding regions well. We can observe that CAR outperformed CRF in removing the over-segmented airways.

\begin{table}
	\centering
	\caption{\textcolor{black}{Comparison of different uncertainty estimation methods used in CAR for the validation set in the first round. $^*$ denotes significant improvement from Initial.}}
	\label{tab:unc_compare}
	\begin{tabular}
		{l|c|c|c} 
		\hline
		& Dice (\%) & RVE (\%)  & HD$_{95}$ (mm) \\ \hline
		Initial & 66.62$\pm$8.53 & 23.91$\pm$15.14 &  11.11$\pm$6.00
		\\ 
		CAR (entropy min) & 67.68$\pm$10.64 & 22.73$\pm$17.38 & 10.54$\pm$6.39
		\\ 
		CAR (Bayesian) & 68.31$\pm$9.52 & 21.80$\pm$16.53 &  10.39$\pm$6.03  
		\\ 
	    CAR (MC Dropout)& \textbf{69.05$\pm$9.41$^*$} & \textbf{21.31$\pm$16.10}  &  \textbf{9.94$\pm$6.16$^*$} \\
		\hline
	\end{tabular}
\end{table}

\textcolor{black}{We also compared MC Dropout~\cite{Gal2016} with two other uncertainty estimation methods: entropy minimization and Bayesian network~\cite{Jena2019}. A visual comparison of them is shown in Fig.~\ref{fig:uncertain_compare}. We found that in spite of longer time required than the other methods, MC Dropout could generate more calibrated uncertainty estimation. As shown in Fig.~\ref{fig:uncertain_compare}, the uncertain region obtained by entropy minimization and Bayesian network~\cite{Jena2019} are mostly located around the border of the segmentation output, and uncertainty map obtained by MC dropout can better indicate under- and over-segmentation regions, which is highlighted by red and yellow arrows in  Fig.~\ref{fig:uncertain_compare}. For quantitative comparison, we applied CAR as a post processing method to the validation set in the first round of I-CRAWL, where these uncertainty estimation methods were used respectively. Results in Table~\ref{tab:unc_compare} show that using MC dropout for CAR improved the prediction accuracy from 66.62\% to 69.05\% in terms of Dice, which outperformed using the other two uncertainty estimation methods. } 

\textcolor{black}{Fig.~\ref{fig:uncertain_rounds} shows a visualization of pseudo labels and uncertainty as the round increases. It can be observed that the initial pseudo label at round 0 has a large under-segmented region with high uncertainty. The pseudo label becomes more accurate and confident as the round increases.  }
\subsubsection{Ablation Study of I-CRAWL}\label{sec:result_ssl_ablation}
For ablation study of I-CRAWL, we set the PF-Net trained only with the annotated images as a baseline, and it was compared with: 1) IT that refers to naive iterative training, where in each round pseudo label of an unannotated image is reset to the prediction given by the network without refinement, 2) IT + CRF that uses standard fully connected CRF~\cite{Krahenbuhl2011} to refine pseudo labels,  3) IT + CAR denoting that our confidence-aware refinement is used to update pseudo labels in each round, and 4) our I-CRAWL (IT + CAR + IW) where IW denotes our confidence-based image weighting of pseudo labels. 

\begin{figure}
	\centering 
	\subfloat[Dice \label{fig:1a}]
	{\includegraphics[width=0.49\linewidth]{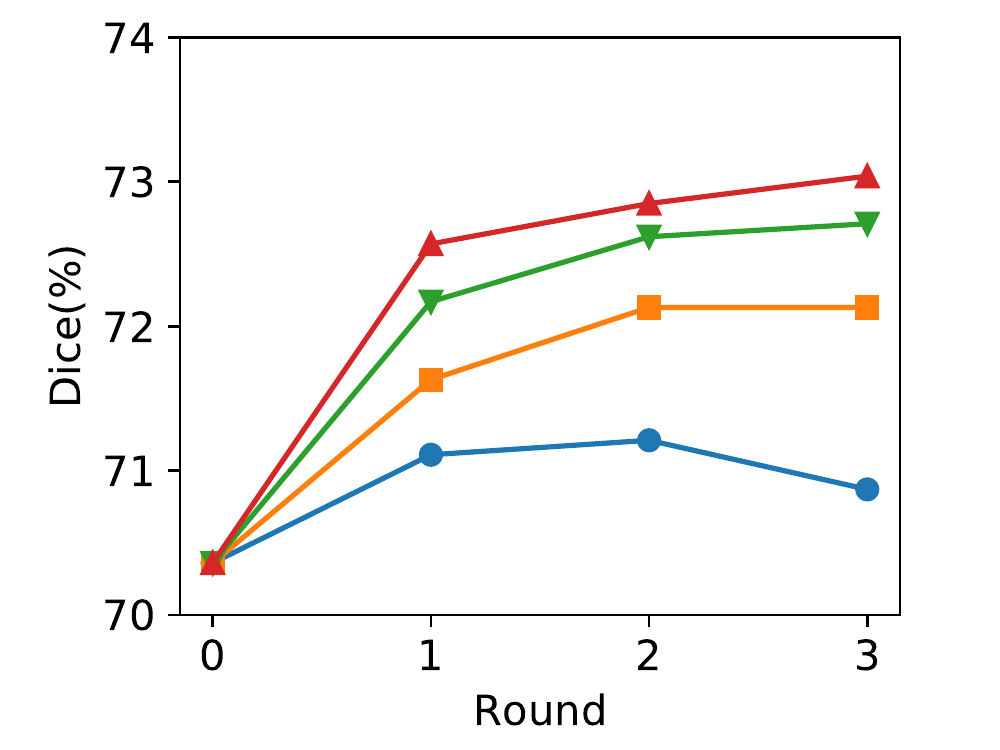}}
	\hfill
	\subfloat[HD$_{95}$ \label{fig:1b}] {\includegraphics[width=0.49\linewidth]{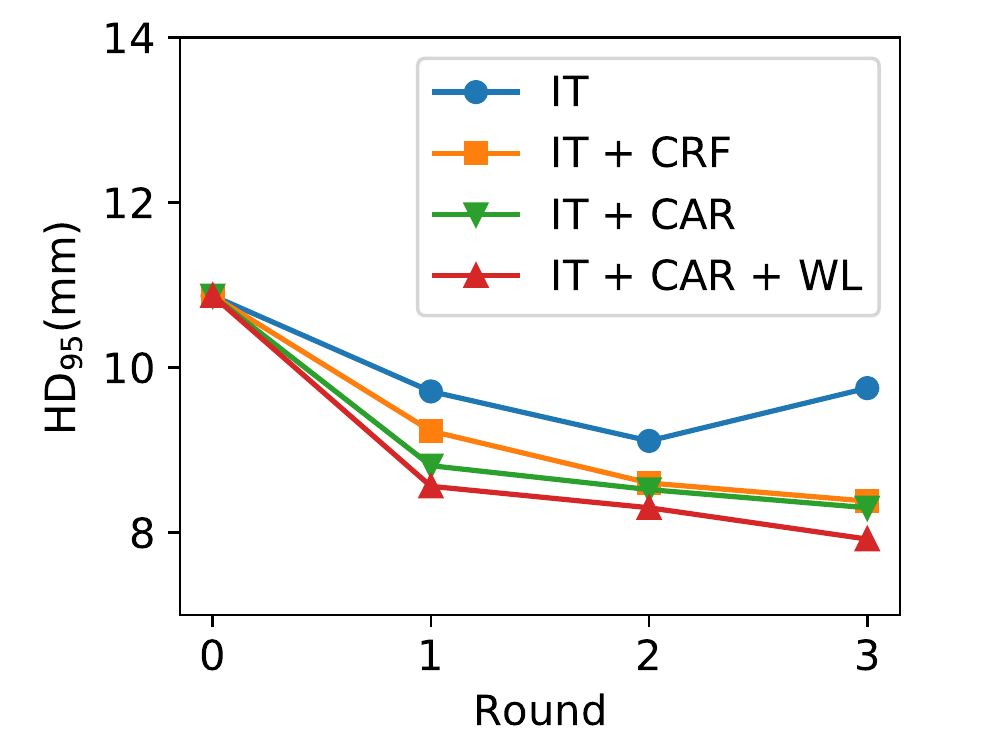}}
	\hfill
	\caption{Comparison between variants of I-CRAWL at different training rounds for semi-supervised learning. Round 0 means the baseline that only learns from annotated images.}
	\label{fig:iteration_num}
\end{figure}

The performance of these methods at different rounds are shown in Fig.~\ref{fig:iteration_num}. Note that round 0 is the baseline, and all the methods based on iterative training performed better than the baseline. However, the improvement obtained by only using IT is slight. Using CRF or CAR to refine the pseudo labels at different rounds achieved a large improvement of Dice, and our CAR considering the voxel-level confidence of predictions outperformed the naive CRF. Weighting of pseudo labels based on image-level confidence  helped to obtain more accurate result, and our I-CRAWL outperformed the other variants. Fig.~\ref{fig:iteration_num} also shows that the improvement from round 0 to round 1 of I-CRAWL is large, but the model's performance does not change much at round 2 and 3. 

Table~\ref{tab:ssl_compare1} shows quantitative comparison between the baseline and variants of I-CRAWL at the end of  training (round 3). It can be observed that IT's performance was not far from the baseline, with an average Dice of 70.87\% compared with 70.36\%. Using CRF and CAR improved the average Dice to 72.13\% and 72.71\%, respectively, showing the superiority of CAR compared with CRF. I-CRAWL improved the average Dice to 73.04\%, with HD$_{95}$ value of 7.92 mm in average, which was better than the other variants.
\begin{table}
	\centering
	\caption{Ablation study of our semi-supervised method I-CRAWL for  PF lesion segmentation. IT: iterative training. CAR: Confidence-aware refinement. IW: Image weighting of pseudo labels. \textcolor{black}{$^*$ denotes significant improvement from the baseline ($p$-value $<$ 0.05).}
	}
	\label{tab:ssl_compare1}
	\begin{tabular}
		{l|c|c|c} 
		\hline
		& Dice (\%) & RVE (\%)  & HD$_{95}$ (mm) \\ \hline
		Baseline & 70.36$\pm$10.14 & 27.96$\pm$21.72 & 10.87$\pm$9.28 
		\\
		IT & 70.87$\pm$8.77 & 28.37$\pm$17.38 & 9.75$\pm$6.36 
		\\  
		IT + CRF & 72.13$\pm$8.97 & 27.38$\pm$20.39 & 8.38$\pm$4.75 
		\\  
		IT + CAR &  72.71$\pm$9.50$^*$ & \textbf{26.05$\pm$18.79} & 8.30$\pm$5.17 
		\\  
		IT + CAR + IW & \textbf{73.04$\pm$10.14$^*$} & 26.12$\pm$20.37   & \textbf{7.92$\pm$4.87$^*$}   \\
		\hline
		\textcolor{black}{Full Supervision} &
		74.58$\pm$10.99 & 23.07$\pm$15.79   & 7.03$\pm$4.09   \\
		\hline
	\end{tabular}
\end{table}

\subsubsection{Comparison with Existing Methods}
 I-CRAWL was compared with several sate-of-the-art semi-supervised methods for medical image segmentation: 1) Fan et al.~\cite{Fan2020} that uses a randomly selected propagation strategy for semi-supervised COVID-19 lung infection segmentation, 2) Bai et al.~\cite{Bai2017a} that uses CRF to refine pseudo labels in an iterative training framework,  corresponding to ``IT + CRF" described previously, 3) Cui et al.~\cite{Cui2019} that is an adapted mean teacher method, and 4) UA-MT~\cite{Yu2019} that is uncertainty-aware mean teacher. For all these methods, we used our PF-Net as the backbone network. 
\begin{table}
	\centering
	\caption{Quantitative comparison of different semi-supervised methods for PF lesion segmentation. \textcolor{black}{$R_{an}$: Ratio of annotated images in the training set. $^\dagger$ denotes there is no significant difference from full supervision ($p$-value $>$ 0.05).}
	}
	\label{tab:ssl_compare2}
	\begin{tabular}
		{l|l|c|c|c} 
		\hline
		$R_{an}$ &Method & Dice (\%) & RVE (\%)  & HD$_{95}$ (mm) \\ \hline
		\multirow{6}{*}{\textcolor{black}{10\%}} & Baseline & 54.73$\pm$16.34 & 45.83$\pm$24.99 & 25.79$\pm$18.71 
		\\
		& Fan et al.~\cite{Fan2020} & 60.44$\pm$15.86 & 42.40$\pm$21.06 & 17.44$\pm$16.00 
		\\  
		& Bai et al.~\cite{Bai2017a} & 65.03$\pm$18.01 & 32.98$\pm$23.16 & 16.82$\pm$14.82 
		\\  
		& Cui et al.~\cite{Cui2019} &  61.86$\pm$16.20 & 41.16$\pm$20.78 & 16.72$\pm$16.18 
		\\  
		& UA-MT~\cite{Yu2019} &  63.94$\pm$15.72 &  38.47$\pm$21.22 &  15.14$\pm$15.34 
		\\  
		& I-CRAWL & \textbf{67.18$\pm$14.02} & \textbf{32.12$\pm$18.50}   & \textbf{11.85$\pm$9.82}   \\ \hline
		\multirow{6}{*}{20\%}&Baseline & 70.36$\pm$10.14 & 27.96$\pm$21.72 & 10.87$\pm$9.28 
		\\
		& Fan et al.~\cite{Fan2020} & 71.62$\pm$10.32 & 27.57$\pm$19.09 & 10.54$\pm$8.41 
		\\  
		& Bai et al.~\cite{Bai2017a} & 72.13$\pm$8.97 & 27.38$\pm$20.39 & 8.38$\pm$4.75 
		\\  
		& Cui et al.~\cite{Cui2019} &  70.39$\pm$10.42 & 26.63$\pm$24.85 & 9.48$\pm$8.82 
		\\  
		& UA-MT~\cite{Yu2019} &  70.46$\pm$9.53 &  27.14$\pm$23.17 &  8.25$\pm$4.93 
		\\  
		& I-CRAWL & \textbf{73.04$\pm$10.14} & \textbf{26.12$\pm$20.37}   & \textbf{7.92$\pm$4.87$^\dagger$}   \\ \hline
		\multirow{6}{*}{\textcolor{black}{50\%}}&Baseline & 72.09$\pm$12.54 & 26.49$\pm$23.19 & 9.38$\pm$6.43 
		\\
		& Fan et al.~\cite{Fan2020} & 72.69$\pm$11.41 & 26.46$\pm$17.71 & 9.19$\pm$6.45 
		\\  
		& Bai et al.~\cite{Bai2017a} & 72.92$\pm$10.41 & 25.71$\pm$19.97$^\dagger$ & 7.45$\pm$4.98$^\dagger$
		\\  
		& Cui et al.~\cite{Cui2019} &  72.72$\pm$12.44 & 24.57$\pm$19.96$^\dagger$ & 8.11$\pm$5.76$^\dagger$ 
		\\  
		& UA-MT~\cite{Yu2019} &  73.40$\pm$11.48 &  24.67$\pm$19.08$^\dagger$ &  7.86$\pm$4.56$^\dagger$
		\\  
		& I-CRAWL & \textbf{74.02$\pm$13.77$^\dagger$} & \textbf{23.93$\pm$16.75$^\dagger$}   & \textbf{7.24$\pm$5.03}$^\dagger$   \\ \hline
		\multicolumn{2}{l|}{\textcolor{black}{Full Supervision}} &
		74.58$\pm$10.99 & 23.07$\pm$ 15.79   & 7.03$\pm$4.09  \\
		\hline
	\end{tabular}
\end{table}

\textcolor{black}{We investigated the performance of  these methods with different ratios of labeled data: 10\%, 20\% and 50\%.} For each setting, the baseline was learning only from the labeled images, \textcolor{black}{and the upper bound was ``full supervision" where 100\% training images were labeled.} The results are shown in Table~\ref{tab:ssl_compare2}. It can be observed that with only 10\% images annotated, the baseline performed poorly with an average Dice of 54.73\%. Our I-CRAWL improved it to 67.18\%, which largely outperformed the other methods. When 20\% images were annotated, our method improved the average Dice from 70.36\% to 73.04\% compared with the baseline, while the mean teacher-based methods did not bring much performance gain. When 50\% images were annotated, our method also outperformed the others, and it was comparable with full supervision (74.02$\pm$13.77\% versus 74.58$\pm$10.99\% in terms of Dice, with $p$-value $>$ 0.05).

\section{Discussion and Conclusion}
\textcolor{black}{Our PF-Net combines 2D and 3D convolutions to deal with anisotropic resolutions, and we set $M=2$ as the in-plane resolution is four times of the through-plane resolution in our dataset. It may be set to other values according to the spacing information of different datasets. Multi-scale guided dense attention in PF-Net is important for dealing with PF lesions with various positions, shapes and scales. We noticed that Sinha et al.~\cite{Sinha2021} also proposed a multi-scale attention, but it has key differences from ours. First, Sinha et al.~\cite{Sinha2021} concatenated feature at different levels of the encoder to obtain a multi-scale feature, which is used as input for parallel attention modules at different scales. While PF-Net learns multi-scale attentions sequentially, where attention at a lower resolution level is used as input for all the higher resolution levels with dense connections.  Second, Sinha et al.~\cite{Sinha2021} used self-attention inspired by non-local block~\cite{Wang2017c} that is computationally expensive with large memory consumption, while PF-Net uses convolution to obtain the attention coefficients at different spatial positions, which has higher memory and computational efficiency. In addition, for each attention module, the attention maps are calculated in two steps in~\cite{Sinha2021}, and a consistency between the two steps is imposed via an L2 distance of their encoded representations, which is called guided attention by the authors. In contrast, guided attention in PF-Net refers to supervising attention maps directly by the resampled segmentation ground truth. }

\textcolor{black}{Our I-CRAWL is a pseudo label-based method for semi-supervised learning. Despite that pseudo label has been previously investigated~\cite{Bai2017a,Fan2020}, I-CRAWL is superior to these works mainly for the following reasons. First, the pseudo labels generated by a model trained with a small set of annotated images inevitably contain a lot of inaccurate predictions. Improving the quality of pseudo labels would benefit the final segmentation model. However, the method in~\cite{Fan2020} does not refine pseudo labels, and Bai et al.~\cite{Bai2017a} refines pseudo labels by CRF without considering their confidence. Our method employs uncertainty estimation to find uncertain regions that are likely to be mis-segmented, and the confidence-aware refinement is more effective to refine these mis-segmented regions, leading to improved accuracy of pseudo labels. Second, the quality of pseudo labels of different images varies a lot, and it is important to exclude low-quality pseudo labels that may corrupt the segmentation model. However, Bai et al.~\cite{Bai2017a} and~\cite{Fan2020} ignored this point and treated all the pseudo labels equally. In contrast, I-CRAWL uses image-level uncertainty information to highlight more confident pseudo labels and down-weight uncertain ones that are unreliable. Thus, the model is less affected by low-quality pseudo labels.}

\textcolor{black}{Uncertainty estimation plays an important role in our I-CRAWL framework. We found that the simple yet effective MC dropout performed better than alternatives including entropy minimization and Bayesian networks~\cite{Jena2019}.  Despite that MC Dropout is slow for uncertainty estimation, it is used offline at the beginning of each round of our method and takes a short time compared with the batch training step.  In the scenario of $20\%$ annotated data,  our uncertainty estimation takes 8.72 minutes (7.69s per 3D image), and CAR and image weighting take 12.98 minutes (11.45s per 3D image). In contrast, the model update with batch training takes around 4 hours for each round, i.e., the first three steps of I-CRAWL account for 8.29\% of the entire runtime of each round, which could be further accelerated by multi-thread parallel computing. Therefore, our uncertainty estimation and CAR require little extra time compared with naive iterative training and Bai et al.~\cite{Bai2017a}. }

In conclusion, we present a novel 2.5D network structure and an uncertainty-based semi-supervised learning method for  automatic segmentation of pulmonary fibrosis from anisotropic CT scans. To deal with complex PF lesions with irregular structures and appearance in CT volumes with anisotropic 3D resolution, we propose PF-Net that combines a 2.5D network baseline with multi-scale guided dense attention. To leverage unannotated images for learning, we propose I-CRAWL that is an iterative training framework, where a confidence-aware refinement process is introduced to update pseudo labels and a confidence-based image weighting is proposed to suppress images with low-quality pseudo labels. Experimental results with lung CT scans from Rhesus Macaques showed that our PF-Net outperformed existing 2D, 3D and 2.5D networks for PF lesion segmentation, and our I-CRAWL could better leverage unannotated images for training than state-of-the-art semi-supervised methods. Our methods can be extended to deal with other structures and human CT scans in the future.


%





\ifCLASSOPTIONcaptionsoff
  \newpage
\fi



\bibliographystyle{IEEEtran}
\bibliography{D:/Documents/texstudio/latex_reference/2020_pf}
\end{document}